\documentclass[]{aa}
\usepackage{subfigure}
\usepackage[varg]{txfonts}                
\usepackage{graphicx}
\usepackage{amssymb}
\usepackage{natbib,twoopt}
\usepackage{comment}
\usepackage[draft]{hyperref}
\bibpunct{(}{)}{;}{a}{}{,} 
\makeatletter
\newcommandtwoopt{\citeads}[3][][]{\href{http://adsabs.harvard.edu/abs/#3}%
{\def\hyper@linkstart##1##2{}%
\let\hyper@linkend\@empty\citealp[#1][#2]{#3}}}
\newcommandtwoopt{\citepads}[3][][]{\href{http://adsabs.harvard.edu/abs/#3}%
{\def\hyper@linkstart##1##2{}%
\let\hyper@linkend\@empty\citep[#1][#2]{#3}}}
\newcommandtwoopt{\citetads}[3][][]{\href{http://adsabs.harvard.edu/abs/#3}%
{\def\hyper@linkstart##1##2{}%
\let\hyper@linkend\@empty\citet[#1][#2]{#3}}}
\newcommandtwoopt{\citeyearads}[3][][]%
{\href{http://adsabs.harvard.edu/abs/#3}
{\def\hyper@linkstart##1##2{}%
\let\hyper@linkend\@empty\citeyear[#1][#2]{#3}}}

\def\beq#1{\begin{equation}\label{#1}}
\def\eeq{\end{equation}}
\def\beqa#1{\begin{eqnarray}\label{#1}}
\def\eeqa{\end{eqnarray}}

\def\Eq#1{Eq.~(\ref{#1})} 

\def\comment#1{\relax}

\usepackage{color}

\begin{document}

\makeatother

\title{Luminosity-dependent changes of the cyclotron line energy and spectral hardness in Cep X-4}

\author{V. Vybornov\inst{1,2}
\and D. Klochkov\inst{1} 
\and M. Gornostaev\inst{3} 
\and K. Postnov\inst{3} 
\and E. Sokolova-Lapa\inst{3}
\and R. Staubert\inst{1} 
\and K. Pottschmidt\inst{4,5} 
\and A. Santangelo\inst{1}}

\offprints{\email{vybornov@astro.uni-tuebingen.de}}

\institute{Institut f\"ur Astronomie und Astrophysik, Universit\"at T\"ubingen, 
Sand 1, 72076 T\"ubingen, Germany
 \and Space Research Institute, 
 Profsouznaya 84/32, Moscow 117997, Russia
\and Sternberg Astronomical Institute, Lomonosov Moscow State University, 
Universitetsky prospekt 13, 119992 Moscow, Russia
\and Department of Physics, University of Maryland Baltimore County, 
Baltimore, MD, 21250, USA
\and CRESST and NASA Goddard Space Flight Center, Astrophysics Science Division, Code 661, 
Greenbelt, MD, 20771, USA}


\abstract{X-ray spectra of accreting pulsars are generally observed to
  vary with their X-ray luminosity. In particular, the hardness of the
  X-ray continuum is found to depend on luminosity.
  In a few sources, the correlation between the energy of
  the cyclotron resonance scattering feature (CRSF) and
  the luminosity is clear. Different types (signs) of the correlation are
  believed to reflect different accretion modes.}
{We analyse two NuSTAR observations of the transient accreting pulsar Cep
  X-4 during its 2014 outburst. Our analysis is focused on a detailed
  investigation of the
  dependence of the CRSF energy and of the spectral hardness on
 X-ray luminosity, especially on short timescales.} 
{To investigate the spectral changes as a function of luminosity within each of the
 two observations, we used the intrinsic variability of the source on
 the timescale of individual pulse cycles (tens of seconds) , 
  the so-called pulse-to-pulse variability.
}
{We find that the NuSTAR spectrum of Cep X-4 contains two CRSFs: the fundamental line at \textasciitilde 30 keV and its harmonic at \textasciitilde 55 keV. We find for the first time that the energy of the fundamental CRSF increases and the continuum becomes harder with increasing X-ray luminosity not only between the two observations, that is, on the long timescale, but also within an individual observation, on the timescale of a few tens of seconds. We investigate these dependencies in detail including their non-linearity. We discuss a possible physical interpretation of the observed behaviour in the frame of a simple one-dimensional model of the polar emitting region with a collisionless shock formed in the infalling plasma near the neutron star surface. With this model, we are able to reproduce the observed variations of the continuum hardness ratio and of the CRSF energy with luminosity.} {}

\maketitle

\section{Introduction}
\label{sec:intro}
Accreting X-ray pulsars are rotating highly magnetised ($B > 10^{12}$~G) neutron stars (NSs) that accrete matter from their normal companion star. 
Close to the NS, within the so-called Alfv\'en radius, magnetic pressure dominates the ram pressure of the ionised accretion flow such that the infalling plasma travels along the magnetic field lines towards the NS magnetic poles,
where it is abruptly decelerated from a free-fall velocity of \textasciitilde 0.6c and heated to $\gtrsim$10$^8$\,K.  
At the NS polar regions, this hot plasma forms structures that
are often called accretion columns, and it radiates in the hard X-ray domain.
To date, more than 200 accreting pulsars are known\footnote{http://www.iasfbo.inaf.it/\textasciitilde mauro/pulsar\_list.html}. 

X-ray spectra of some accreting pulsars exhibit absorption-line-like features called cyclotron resonance scattering features (CRSFs), or cyclotron lines. The first such line was discovered in 1976 during a balloon observation of Her X-1 \citepads{1978ApJ...219L.105T}. 
The cyclotron feature permits a direct measurement of the neutron star magnetic field strength because the latter is directly related to the cyclotron line centroid energy $E_{\textrm{CRSF}}$ as $E_{\textrm{CRSF}} = 11.6[\textrm{keV}]B_{12}n/(z+1)$, where $B_{12}$ is the magnetic field strength in units of $10^{12}$ Gauss, $n$ is the Landau level number, and $z$ is the gravitational redshift at the line-forming region. The fundamental cyclotron line corresponds to $n = 1$. Cyclotron lines corresponding to higher $n$ are referred to as harmonics.

The cyclotron feature has been found to change with the X-ray luminosity $L_\text{X}$ in the spectra of a few accreting pulsars. 
The correlation between the CRSF centroid energy $E_\text{CRSF}$ and $L_\text{X}$ has so far been firmly confirmed in six sources.
Four of them exhibit a positive correlation for the fundamental line, one for the first harmonic, and the last pulsar exhibits a negative correlation for the fundamental line. 
Below we provide a brief summary of the $E_\text{CRSF}/L_\text{X}$ correlations observed in these sources.
The correlation in Cep X-4 is described separately.
\begin{itemize}
\item \textit{Her X-1} is the first source where a positive $E_\text{CRSF}/L_\text{X}$
correlation was reported by \citetads{2007A&A...465L..25S} using 
RXTE observations. \citetads{2011A&A...532A.126K} confirmed the correlation on a short timescale 
using a pulse-to-pulse analysis technique, which is also used in our work. Her X-1 demonstrates another phenomenon, however. The energy of the CRSF in this source apparently changes with time \citepads[][and references therein]{2016A&A...590A..91S}. This is the first source for which a secular change of the CRSF was reported.
\item \textit{GX 304-1}. \citetads{2011PASJ...63S.751Y} discovered a CRSF during the 2010 outburst and reported a marginal indication of a positive correlation between its energy and the X-ray flux. 
The correlation was observed with high significance using INTEGRAL \citepads{2012A&A...542L..28K, 2015A&A...581A.121M} and RXTE \citepads{2016arXiv161008944R} data. 
\item \textit{A 0535+26}. The source spectrum has two CRSFs: the fundamental and the first harmonic \citepads{1994A&A...291L..31K, 2008A&A...480L..17C}. A positive $E_\text{CRSF}/L_\text{X}$ correlation for the fundamental line was first found on a short timescale, using the pulse-to-pulse technique applied to RXTE and INTEGRAL/ISGRI observations of a source outburst \citepads{2011A&A...532A.126K}. 
The correlation was subsequently confirmed on longer timescales by \citetads{2015ApJ...806..193S} using INTEGRAL/SPI data.
\item \textit{Vela X-1}. The source spectrum also has two CRSFs interpreted as the fundamental line and the first harmonic, but the fundamental feature is considerably weaker than the first harmonic. A positive $E_\text{CRSF}/L_\text{X}$ correlation for the harmonic line is observed in the source spectrum \citepads{2014ApJ...780..133F, 2016MNRAS.463..185L}.
\item \textit{V 0332+53}. The fundamental cyclotron line in the source spectrum was discovered during the 1983 outburst with \textit{Tenma} \citepads{1990ApJ...365L..59M}. During subsequent outbursts, two harmonics were detected \citepads{2005ApJ...634L..97P, 2005A&A...433L..45K}. The energy of the fundamental CRSF is negatively correlated with the source X-ray flux \citepads{2006A&A...451..187M, 2010MNRAS.401.1628T, 2016arXiv160703933D}. This is the only source for which a negative $E_\text{CRSF}/L_\text{X}$ correlation is confirmed so far. There are indications that the source, similarly to Her X-1, might exhibit another type of variability: \citetads{2016MNRAS.460L..99C} reported a change in the energy of the fundamental CRSF at the same level of luminosity.
\end{itemize}
There are also X-ray pulsars in which an indication for a correlation between the CRSF energy and luminosity is reported but so far not unambiguously confirmed, such as 4U 0115+63 \citepads{2013A&A...551A...6M, 2015MNRAS.454..741I} and 4U 1538-522 \citepads{2014ApJ...792...14H}.  

Here, we focus on the accreting pulsar Cep X-4 discovered in 1972 with the Orbiting Solar Observatory-7 
\citepads{1973ApJ...184L.117U} and re-discovered in 1988 with \textit{Ginga} as GS 2138+56 during a one-month-long X-ray outburst  \citepads{1988IAUC.4575....1M}. 
Several other outbursts have occurred since the discovery. 
Pulsations with a period of 66.2 s were detected \citepads{1991ApJ...366L..19K} during the 1988 outburst, which established the nature of the source as an accreting magnetised NS. 
The CRSF was discovered at around 30 keV during the same outburst \citepads{1991ApJ...379L..61M}. 
A Be star was proposed as an optical counterpart during observations of the 1993 outburst 
\citepads{1997IAUC.6698....2R}. 
This identification was confirmed later by \citetads{1998A&A...332L...9B}, who also estimated the distance to the source to be $3.8\pm0.6$ kpc. 
During the 2002 outburst, the CRSF energy was re-measured with RXTE to be at $30.7_{-1.9}^{+1.8}$ keV without a considerable dependence on X-ray luminosity, although the continuum was reported
to become harder with increasing luminosity \citepads{2007A&A...470.1065M}. The source went into outburst again in 2014 and was observed with the Nuclear Spectroscopic Telescope Array \citepads[NuSTAR,]{2013ApJ...770..103H} by the two detectors FPMA and FPMB. Two NuSTAR \textcolor[rgb]{1,0.4,0.4}{}observations of Cep X-4 were performed during this outburst (Fig.~\ref{fig:BAT_NuSTAR}). The averaged fluxes in these observations differ by a factor of about 3.5, providing an opportunity to investigate the spectrum-luminosity dependence. 
The CRSF energy was found to be different between the two observations: the lower cyclotron line energy corresponds to the fainter observation, indicating a possible positive correlation between the CRSF energy and luminosity, although with only two data points \citepads{2015ApJ...806L..24F}.

In our work, we consider the behaviour of the CRSF energy and of the continuum hardness ratio as a function of X-ray luminosity within individual NuSTAR observations of the 2014 outburst using the pulse-to-pulse variability of the source. The structure of the paper is as follows. In Sect. 2 we present a summary of the observations and data reduction. In Sect. 3 we describe our analysis of the pulse-phase averaged spectra and of the pulse-amplitude resolved spectra. Theoretical implications of the results are discussed in Sect. 4.
Section 5 provides a summary and conclusions.

\section{Observations and data reduction}
\label{sec:obs}
\begin{figure}[h]
\center{\includegraphics[width=1\linewidth]{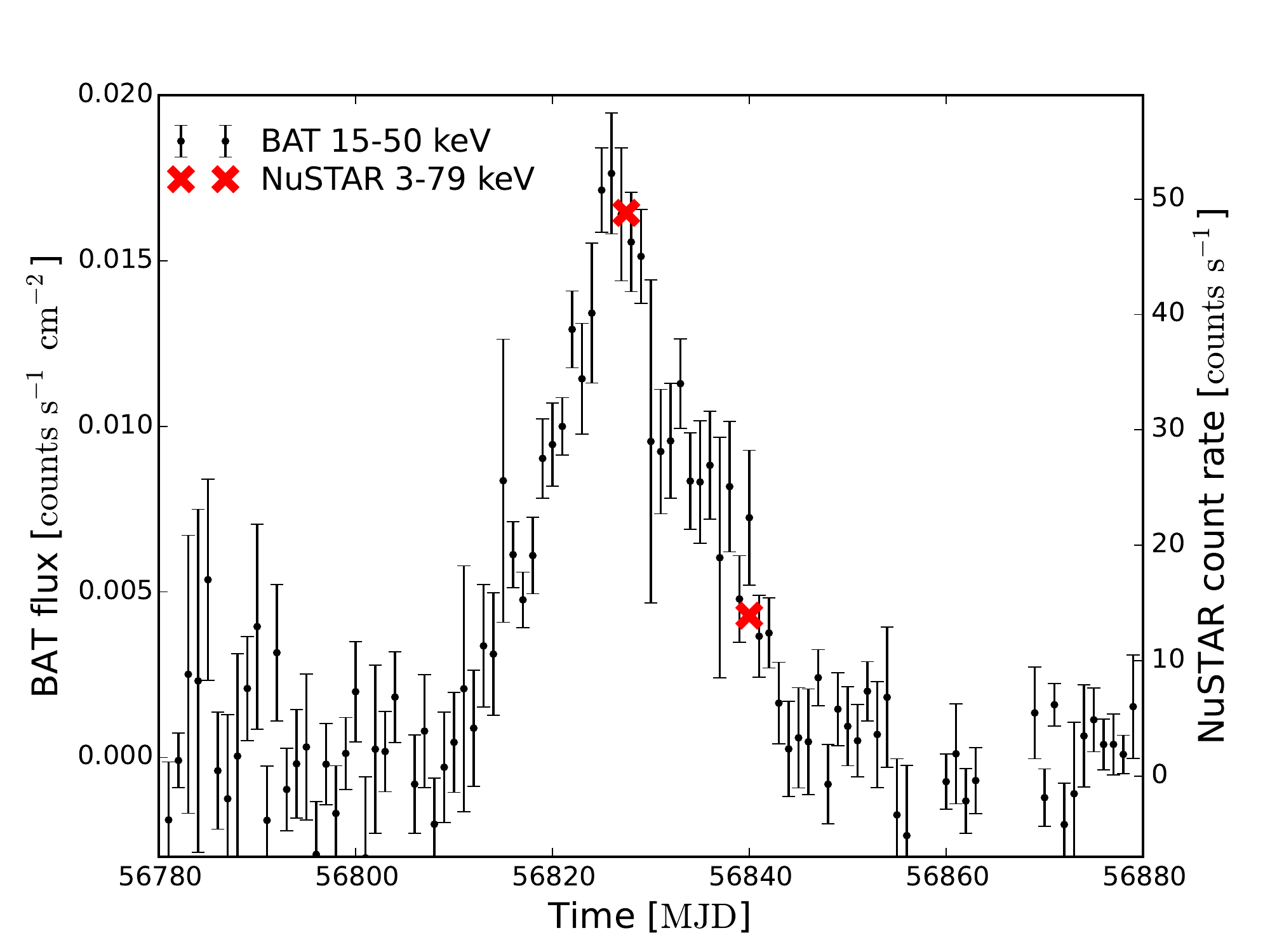}}
\caption{Light curve of the 2014 outburst of Cep X-4 taken with Swift/BAT (black data points). The NuSTAR observations analysed here are indicated with red crosses.}
\label{fig:BAT_NuSTAR}
\end{figure}

NuSTAR observed the source twice during the 2014 outburst, on June 18-19 and on July 1-2 (Fig.~\ref{fig:BAT_NuSTAR}). The first observation is near the outburst maximum with a total on-source time of 47.9 ks (ObsID 80002016002, MJD 56826.92-56827.84). It is referred to as observation~I in the following. The second observation, referred to as observation~II, was performed at the decay of the outburst with a total on-source time of 45.2 ks (ObsID 80002016004, MJD 56839.43-56840.31). Figure~\ref{fig:BAT_NuSTAR} also represents the Swift/BAT observations of the outburst\footnote{http://swift.gsfc.nasa.gov/results/transients/weak/Ginga2138p56/}.

In our pulse-to-pulse analysis described in Sect.~3.2, all source spectra of the two NuSTAR detectors FPMA and FPMB are extracted from a circular region with a radius of 80'' on the detector plane.
The background spectra are extracted from a circular region of the same radius at the periphery of the NuSTAR field of view, similarly to \citetads{2015ApJ...806L..24F}. 

The same extraction regions were initially also used for our pulse-averaged spectra. For a detailed investigation of the possible cyclotron harmonic feature around 55 keV, we used modified extraction regions as described in Sect.~3.1, however.

The data processing was performed using the standard \texttt{nupipeline} and \texttt{nuproducts} utilities and XSPEC\footnote{https://heasarc.gsfc.nasa.gov/xanadu/xspec/} v12.9 as part of HEASOFT\footnote{http://heasarc.nasa.gov/lheasoft/} 6.18.

\section{Spectral analysis and results}
\label{sec:spe}
\subsection{Pulse-averaged spectrum}
\begin{figure}[h]
\subfigure[]{
\includegraphics[width=1\linewidth]{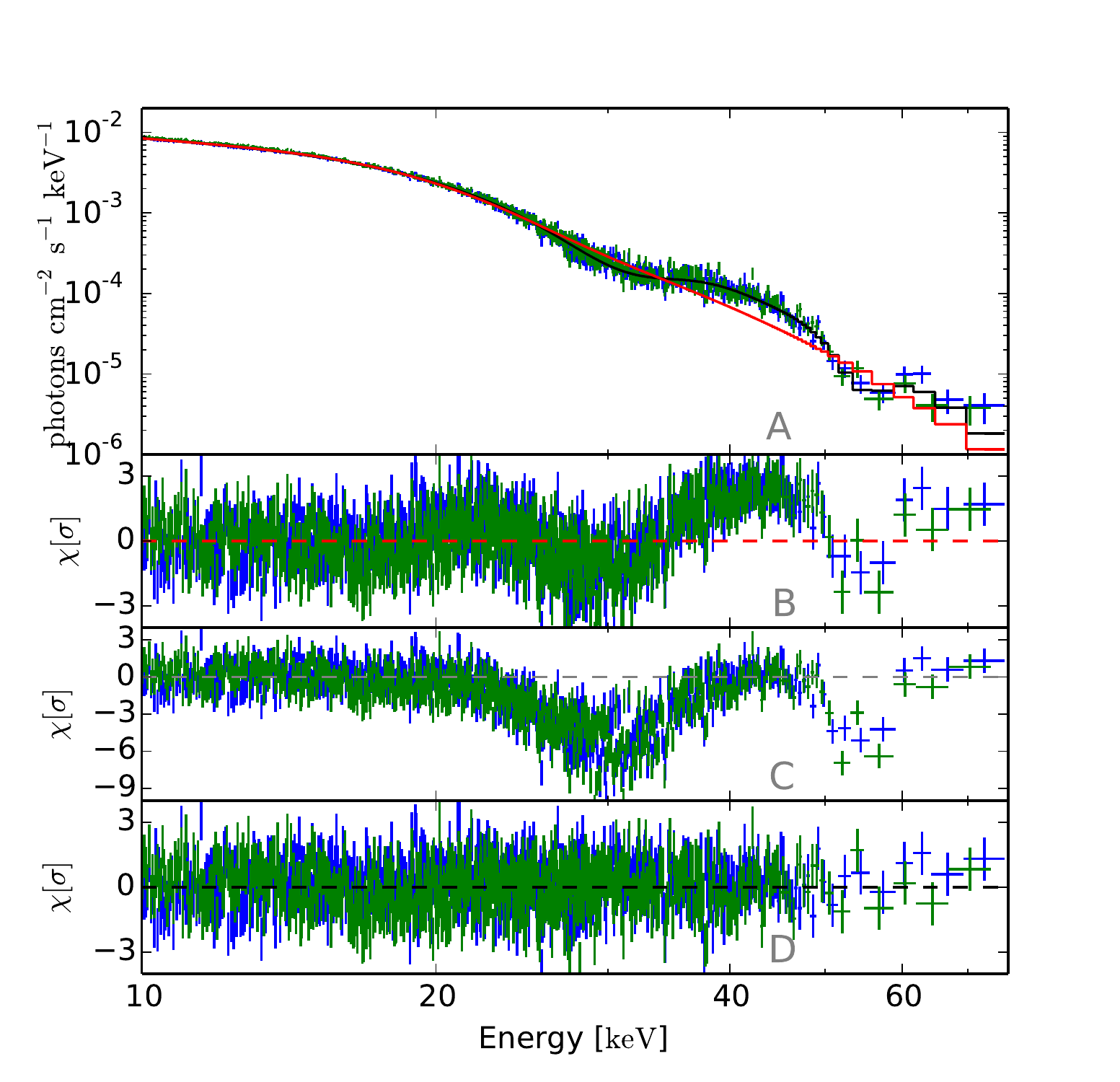}
\label{fig:spec_002}
}
\vfill
\subfigure[]{
\includegraphics[width=1\linewidth]{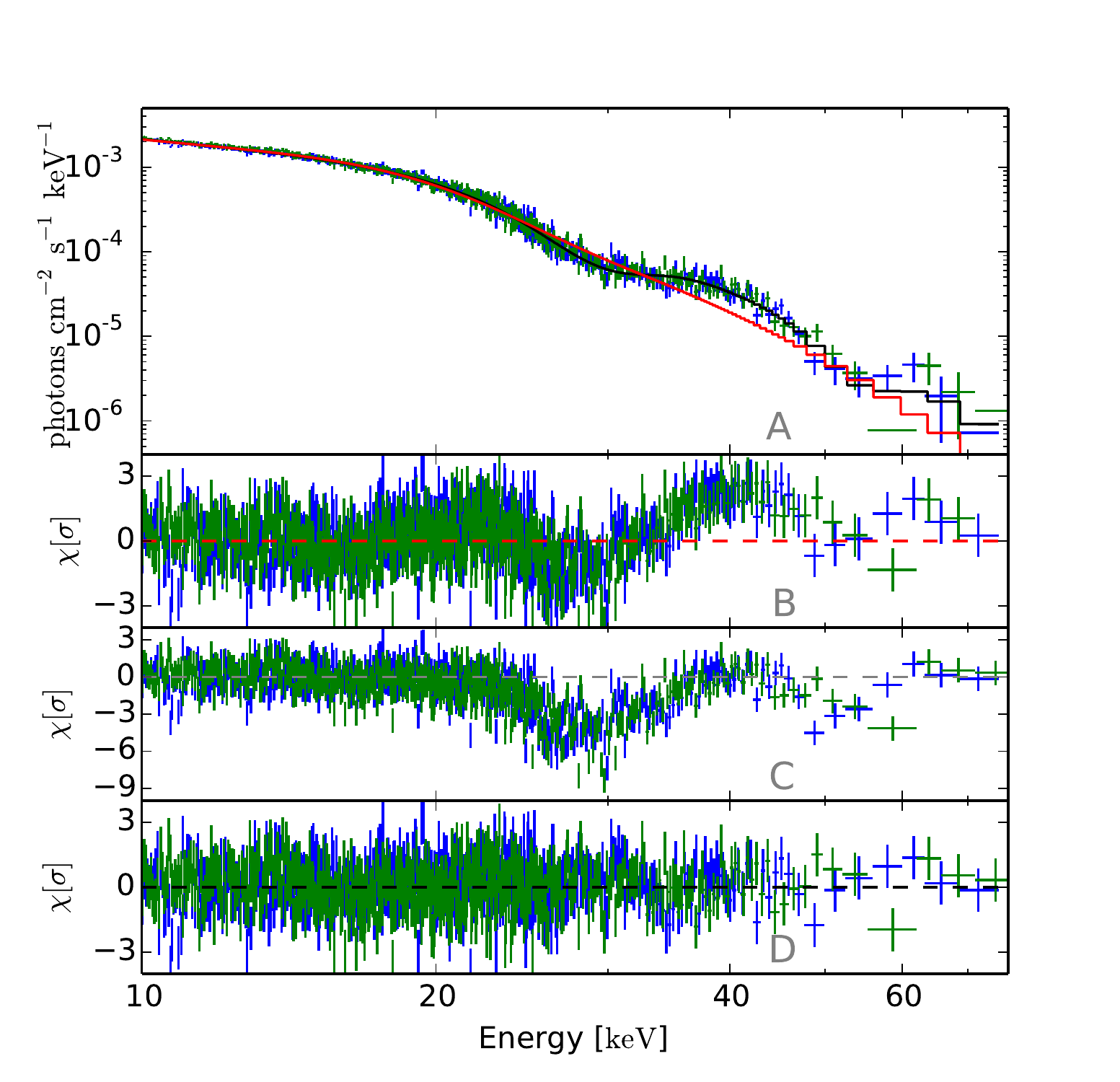}
\label{fig:spec_004}
}
\caption{Pulse-averaged NuSTAR spectra of obs.~I (a) and obs.~II (b). The blue and green data points correspond to FPMA and FPMB, respectively. Panel~B shows residuals to the model without cyclotron lines (red line). 
Panel~C shows residuals to the model setting the strengths of the the
two cyclotron lines to zero, but leaving the other parameters at their
best-fit values obtained using the model with the lines 
(in order to illustrate the contribution of the lines to the continuum). Residuals to the best-fit model (black line in panel~A) are shown in panel~D.
The best-fit parameters of the model with the lines are listed in Table 1.}
\label{fig:specs}
\end{figure}
The spectral continuum of highly magnetised accreting neutron stars is typically described with phenomenological power law models with an exponential cutoff at higher energies \citepads[e.g.][]{2002ApJ...580..394C}. Some models, however, include a detailed description of the physics of the polar emitting region \citepads{1981A&A....93..255W, 2007ApJ...654..435B, 2012A&A...538A..67F, 2014A&A...562A..99C}. These physical models have been successfully applied in some cases \citepads{2009A&A...498..825F,  2016A&A...591A..29F, 2016arXiv160808978W}. However, assumptions and approaches used in these models are still being tested and revised.
Since our work is focused on the behaviour of the CRSF centroid energy and of the general properties of the continuum, like its hardness, it is sufficient to describe the spectra using the traditional phenomenological models, which also allows us to compare our results with the previous works.  

\citetads{2007A&A...470.1065M} showed that the continuum of the Cep X-4 pulse-averaged spectrum between 3.5 and 70 keV is described well by a power-law function (\texttt{powlaw} in XSPEC) with a Fermi-Dirac cut-off \citepads[\texttt{fdcut}, ][]{1986LNP...255..198T} 
\begin{equation}
\textrm{fdcut(E)} \propto E^{-\Gamma}\frac{1}{1 + \exp\left(\frac{E-E_{\textrm{cut}}}{E_{\textrm{fold}}}\right)}
,\end{equation}
with the inclusion of a wide additive Gaussian emission line (\texttt{gauss} in XSPEC) at around 14 keV required to account for a bump-like feature in the residuals between 8 and 20 keV. 

A broad line-like feature like this in the residuals between these energies, known as the 10~keV feature, is reported for some other accreting pulsars observed with RXTE, \textit{Ginga}, BeppoSAX, and other missions \citepads{2002ApJ...580..394C}.
To flatten the 10~keV feature in the Cep X-4 spectrum, we can use not only a broad emission line, but also an absorption line or a combination of the two. The additional wide Gaussian emission line does not correspond to a separate physical component. The 10~keV feature thus apparently reflects limitations of the simple phenomenological function used to describe the continuum of the high-quality NuSTAR spectrum of Cep X-4.

We describe the fundamental CRSF, clearly seen in the NuSTAR spectra at around 30 keV (Fig.~\ref{fig:specs}), as a multiplicative absorption line with a Gaussian optical depth profile (\texttt{gabs} in XSPEC). The Fe $K_\alpha$ fluorescent line at around 6.4 keV is modelled as an additive Gaussian emission line. We also take the photoelectric absorption into account in the soft part of the spectra using the \texttt{phabs} model in XSPEC with photoelectric cross-sections from \citetads{1996ApJ...465..487V} and element abundances from \citetads{2000ApJ...542..914W}.

\begin{table}
\caption{Parameters of the best-fit models (see Sect. 3.1) of the pulse-averaged spectra represented in Fig.~\ref{fig:specs} in the energy range of 3-79 keV from a circular region with a radius of 60''. }
\label{table:fit_params}
\centering
\begin{tabular}{l c c}
\hline\hline
Parameter & Observation I & Observation II \\
\hline
$N_H \, [10^{22} \, \textrm{cm}^{-2}]$ & $2.6_{-0.3}^{+0.3}$ & $2.8_{-0.4}^{+0.3}$ \\
$\Gamma$ & $1.29_{-0.06}^{+0.07}$ & $1.33_{-0.06}^{+0.06}$ \\
$A\tablefootmark{a}$ & $0.21_{-0.03}^{+0.04}$ & $6.6_{-0.6}^{+0.7} \times 10^{-2}$ \\
$E_{\textrm{cut}}$ [keV] & $13_{-6}^{+4}$ & $13 \,\textrm{(fixed)}$ \\
$E_{\textrm{fold}}$ [keV] & $10.3_{-0.9}^{+1.0}$ & $10.5_{-0.5}^{+0.5}$ \\
$E_{\textrm{add}}\tablefootmark{b}$ [keV] & $11.5_{-0.4}^{+0.4}$ & $12.9_{-0.5}^{+0.5}$  \\
$\sigma_{\textrm{add}}$ [keV] & $5.9_{-0.3}^{+0.3}$ & $5.7_{-0.4}^{+0.4}$ \\
$A_{\textrm{add}}\tablefootmark{a}$ & $4.5_{-0.7}^{+0.7} \times 10^{-2}$ & $7.3_{-1.2}^{+1.3} \times 10^{-3}$ \\
$E_{\textrm{Fe}\,K_\alpha}$ [keV] & $6.53_{-0.02}^{+0.02}$ & $6.42_{-0.03}^{+0.03}$ \\ 
$\sigma_{\textrm{Fe}\,K_\alpha}$ [keV] & $0.35_{-0.03}^{+0.04}$ & $0.22_{-0.05}^{+0.05}$  \\
$A_{\textrm{Fe}}\tablefootmark{a}$ & $1.04_{-0.07}^{+0.08} \times 10^{-3}$ & $1.9_{-0.3}^{+0.3} \times 10^{-4}$ \\
$E_{\textrm{CRSF}}$ [keV] & $30.6_{-0.3}^{+0.2}$ & $29.0_{-0.2}^{+0.3}$ \\
$\sigma_{\textrm{CRSF}}$ [keV] & $3.9_{-0.2}^{+0.2}$ & $3.8_{-0.2}^{+0.2}$ \\
$d_{\textrm{CRSF}}\tablefootmark{c}$ & $6.6_{-0.8}^{+0.9}$ & $7.0_{-0.6}^{+0.8}$ \\
$E^{\textrm{h}}_{\textrm{CRSF}}\tablefootmark{d}$ [keV] & $54.8_{-0.5}^{+0.5}$ & $54.0_{-1.4}^{+1.5}$ \\
$\sigma^{\textrm{h}}_{\textrm{CRSF}}$ [keV] & $3.2_{-0.4}^{+0.5}$ & $4.4_{-0.9}^{+1.1}$ \\
$d^{\textrm{h}}_{\textrm{CRSF}}$ & $9.6_{-1.8}^{+2.0}$ & $9.2_{-2.9}^{+3.3}$ \\
$C_{\textrm{FPMB}}\tablefootmark{e}$ & $1.0374 \pm 0.0013$ & $1.024 \pm 0.002$ \\
$\textrm{Flux}_ {3-78 \, \textrm{keV}}\tablefootmark{f}$& $26.9 \pm 0.1$ & $7.10 \pm 0.02$ \\
\hline
$\chi^2_{\textrm{red}}/\textrm{d.o.f.}$ & 1.06/1611 & 1.05/1320 \\
\hline
\end{tabular}
\tablefoot{
The indicated uncertainties are at $1\sigma$ (68\%) confidence level. 
\tablefoottext{a}{In photons $\textrm{s}^{-1} \, \textrm{cm}^{-2} \, \textrm{keV}^{-1}$ at 1 keV.}\\
\tablefoottext{b}{The parameters of the additional wide Gaussian emission line are denoted by the "add" subscript.} \\
\tablefoottext{c}{Optical depth $\tau = d/(\sigma \sqrt{2\pi})$.} \\
\tablefoottext{d}{The "h" superscript marks the harmonic line parameters.}\\
\tablefoottext{e}{A cross-calibration constant for FPMB relative to FPMA.}\\
\tablefoottext{f}{Unabsorbed flux in  $10^{-10} \, \textrm{erg} \, \textrm{s}^{-1} \, \textrm{cm}^{-2}$.}
}
\end{table}
 
Panels B of Fig.~\ref{fig:specs} show sharp residuals around 55 keV. These residuals are present in the spectra of the the
two NuSTAR observations in the data of FPMA and FPMB. To describe this line-like feature as well as possible and to clarify whether the feature could be a harmonic of the CRSF, we analysed the spectrum in the energy range between 50 and 70 keV. 
First, we used background spectra extracted from different areas on the detector plane and found that the feature does not disappear. The dependence of the effective area on energy does not contain any peculiarities around 55 keV either. Second, we performed an analysis searching for an optimal size of the source extraction region at which the signal-to-noise ratio has a maximal value in the energy range of 50-70 keV. We found that the optimal size lies between 50" and 70" and extracted the source spectrum from a 60" radius region. Background spectra were extracted from a circular region of 160" radius at the periphery of the NuSTAR field of view.
This adjustment of the extraction region does not significantly
change the best-fit parameters of the continuum and of the fundamental cyclotron line.
We then built the source spectrum describing the feature with the \texttt{gabs} model (Fig.~\ref{fig:specs}). 
A very similar feature is present in the Suzaku spectrum of Cep X-4 taken during observations of the same outburst. \citetads{2015MNRAS.453L..21J} analysed the Suzaku data and interpreted the feature as the first harmonic. We infer from the above that the feature at around 55 keV in the Cep X-4 spectrum is indeed the first harmonic of the CRSF.

Panel C of Fig. 2 shows the contribution of the lines to the continuum. Both lines are well localised and do not dominate the continuum, in contrast to the case of 4U$\,$0115+63, for
example, where inadequately modelled cyclotron lines could fully distort the continuum, as illustrated in Fig.~8 of \citetads{2013A&A...551A...6M}
for instance.

The applied model can therefore be written as
\begin{equation}
\textrm{ABS} \times (\textrm{FDCUT} \times \textrm{CRSF}\times \textrm{CRSF}_h + \textrm{ADD} + \textrm{Fe}),
\end{equation}
where ABS is the photoelectric absorption (\texttt{phabs}), FDCUT is the \texttt{fdcut} function, ADD is the additional emission Gaussian line in the continuum, the $\textrm{CRSF}$ and $\textrm{CRSF}_h$ are \texttt{gabs} models for the fundamental CRSF and the harmonic, respectively, and Fe is the iron $K_\alpha$ complex modelled as a Gaussian emission line.
The best-fit parameters of this model for the pulse-phase averaged spectra for the two NuSTAR observations are listed in Table~1.
Owing to lower statistics, the spectra of observation~II do not allow us to constrain $E_{cut}$. Since we prefer to use the same spectral model in both observations, we fixed $E_{cut}$ in observation~II to the best-fit value we obtained in observation~I.

\subsection{Pulse-to-pulse analysis}

To investigate the variation of spectral parameters as a function of X-ray luminosity within individual observations, we applied the pulse-to-pulse technique elaborated in \citetads{2011A&A...532A.126K} and \citetads{2013A&A...552A..81M}. This method is aimed at investigating the spectrum-luminosity dependence on short timescales using the pulse-to-pulse variability that is generally exhibited by accreting pulsars \citepads{1980ApJ...239.1010S, 1985ApJ...298..585F, 2007AstL...33..368T}.

\begin{figure}[h]
\center{\includegraphics[width=1\linewidth]{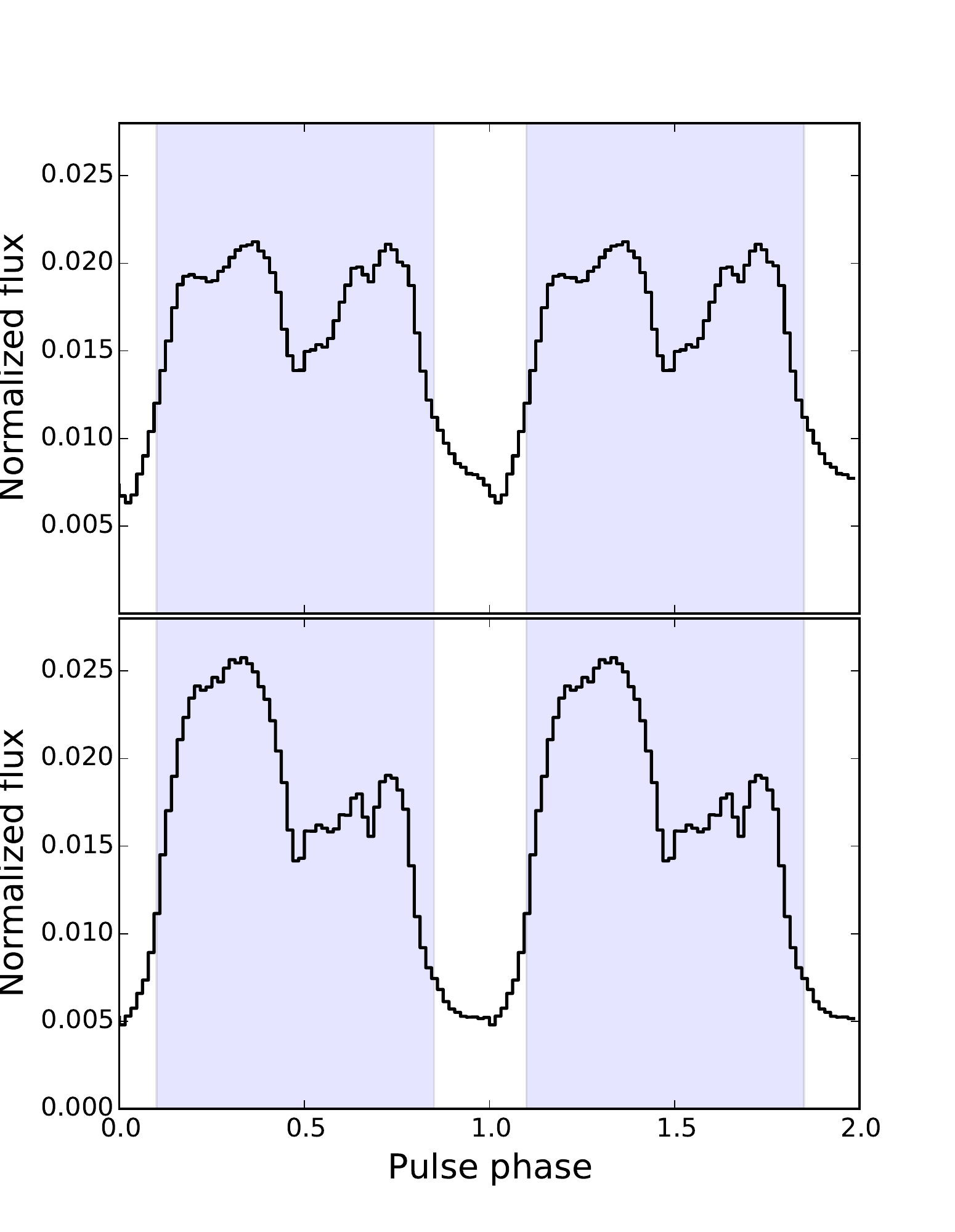}}
\caption{NuSTAR pulse profiles in the 3-79 keV energy range from observations~I (top) and~II (bottom). The shaded areas indicate the phase interval considered for the pulse-to-pulse analysis. The part of the profile within the interval is referred to as \textit{pulse} (see text).}
\label{fig:pulse_profs}
\end{figure}

\begin{figure}[h]
\center{\includegraphics[width=1\linewidth]{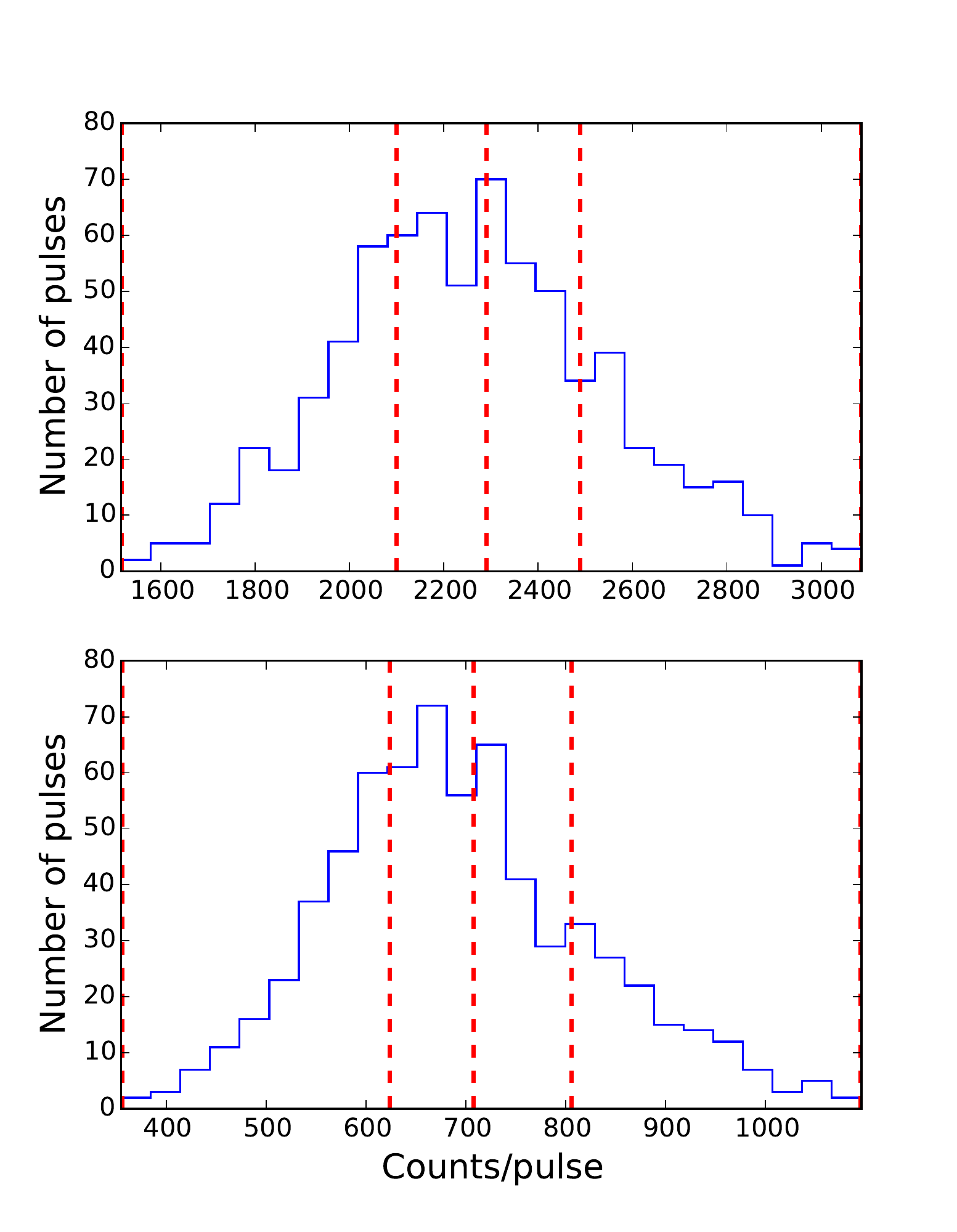}}
\caption{Distributions of counts in individual pulse (i.e. of pulse amplitudes) for observations~I (top) and~II (bottom). The red dashed lines indicate the boundaries of the amplitude bins we used to extract amplitude-resolved spectra. In each distribution, the total number of counts is evenly distributed among the bins.}
\label{fig:pulse_distribs}
\end{figure}

Our pulse-to-pulse analysis includes the following stages. 
First, all photon arrival times are barycentered using the DE-200 solar system ephemeris. 
We measured pulse periods of 66.3352 $\pm$ 0.0003 and of 66.3336 $\pm$ 0.0002 s for observation~I and observation II, respectively. 
These periods are used to accumulate the pulse profiles shown in Fig.~\ref{fig:pulse_profs}.
We then specify a phase interval containing the brightest part of the pulse profile, which is referred to as \textit{pulse} (shaded areas in Fig.~\ref{fig:pulse_profs}). 
To characterise the brightness of the pulse, we define the \textit{pulse amplitude} as the total number of counts in the entire NuSTAR energy band (3--79\,keV) within the specified phase interval of an individual pulsation cycle divided by the width of this interval.
The number of counts in a single pulse is insufficient to extract a meaningful spectrum, therefore we group pulses with similar amplitudes together, similarly to \citetads{2011A&A...532A.126K}. 
For this purpose, we construct the distribution of pulses over amplitude, which is shown in Fig.~\ref{fig:pulse_distribs}.
Although the average flux within each observation does not change much, the distribution clearly covers a dynamical range of a factor of two in amplitude, indicating a strong intrinsic pulse-to-pulse variability of the source. 
We then split the distribution into four bins and keep approximately the same number of counts in each
bin (the dashed vertical lines in Fig.~\ref{fig:pulse_distribs}). The width of the bins is wider than the statistical fluctuation (variance) of pulse amplitudes, which guarantees that the observed variability reflects the intrinsic flux changes. 
Finally, we extract a broadband
spectrum for each pulse amplitude bin . 

Since the number of photons in each pulse amplitude bin spectrum is lower than 25\% of the pulse-averaged spectrum, the harmonic becomes insignificant, especially in observation II. Since our purpose in this part is to investigate the variability of the energy of the fundamental CRSF, we extracted spectra between 3 and 48 keV to eliminate a possible influence of the residuals around the harmonic energy on the fundamental line. Therefore
we here use the same model as in the analysis of the pulse-averaged spectrum described in Sect. 3.1, except for the harmonic component, and we do not fix the cut-off $E_{cut}$ energy in observation II.

\begin{figure}[h]
\center{\includegraphics[width=1\linewidth]{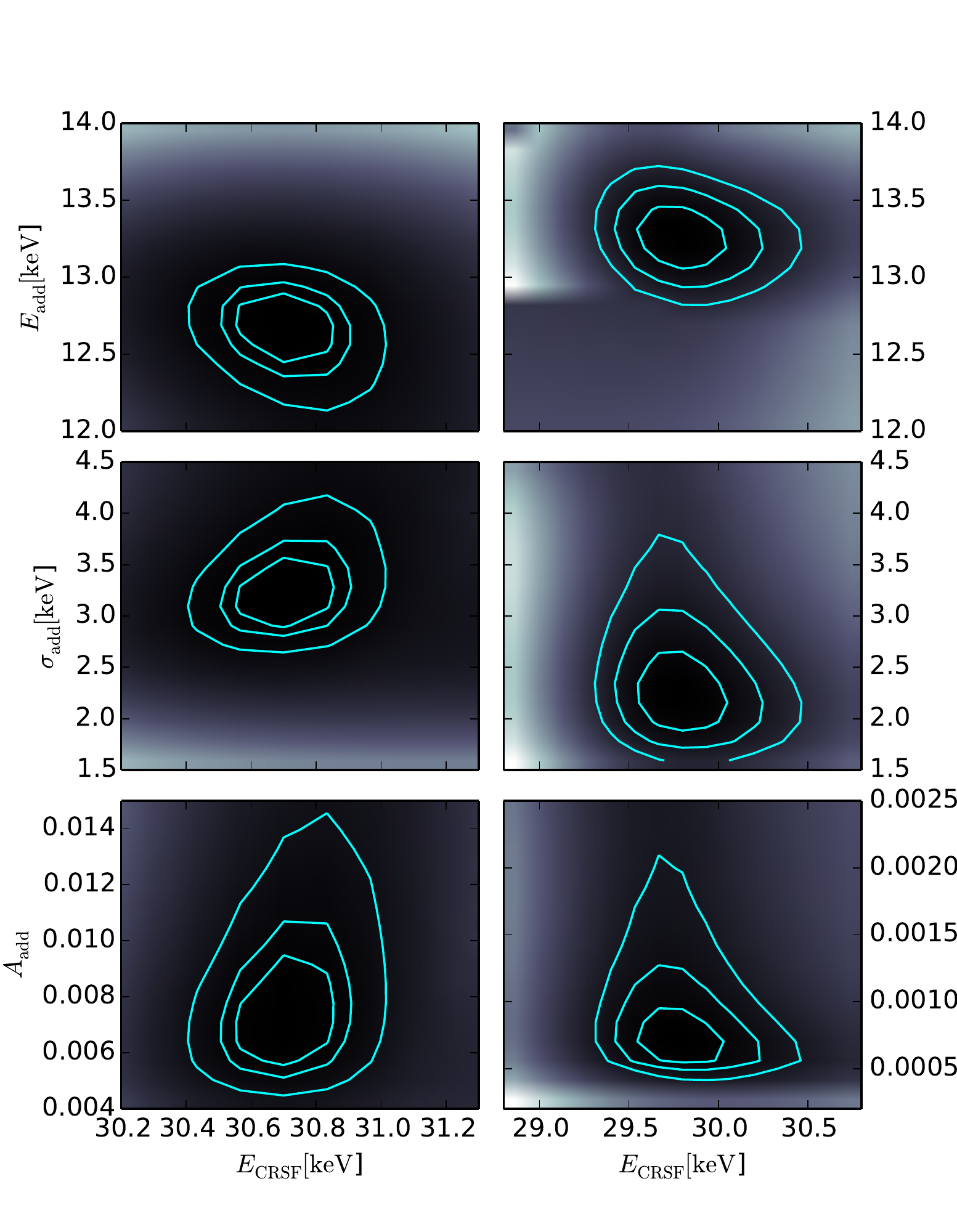}}
\caption{Confidence regions to investigate possible intrinsic
    correlation between the CRSF centroid energy $E_{CRSF}$
    and the parameters of the Gaussian line describing the 10~keV
    feature in the continuum. The left and right columns show results
    for observations~I and~II, respectively. The contours
    correspond to the 1$\sigma$, 2$\sigma,$ and 3$\sigma$ confidence
    levels for the two parameters of interest. The colour scale indicates
    $\Delta\chi^2$: darker areas correspond to 
    lower $\Delta\chi^2$. $A_\mathrm{add}$ is in
    photons $\textrm{s}^{-1} \, \textrm{cm}^{-2} \, \textrm{keV}^{-1}$
    at 1 keV.}
\label{fig:conts}
\end{figure}

As mentioned above, to describe the continuum successfully, we
  introduced an additional component to flatten the 10~keV
  feature. To ensure that this component does not influence the CRSF
  centroid energy \citepads[see e.g.][]{2013AstL...39..375B}, 
  we constructed the contour plots for the CRSF energy versus
  the parameters of the 10~keV component
  (Fig.~\ref{fig:conts}). The confidence regions in the plots do not
  indicate any significant correlation between the parameters. We  therefore conclude that the reported variations in the line energy
  are not associated with possible variations in the 10~keV feature.

From a circular region with a radius of 80'' (see Sect.~2), we extracted four spectra per detector for each observation, corresponding to the four pulse amplitude bins. 
In our fits, we left the relative normalisation of the two detectors free because there is a slight difference in their calibration.
The model parameters of the spectra of the amplitude bins were coupled except for the parameters whose behaviour as a function of pulse amplitude is investigated.
The investigated parameters are the energy of the fundamental cyclotron line $E_\text{CRSF}$ and the power-law index $\Gamma$. The continuum hardness HR considered below is defined as the
ratio of the flux in the 8-12 keV range to the flux in the 4-6 keV range. All other parameters (except for the relative normalisation, as mentioned above) were coupled.
We verified that the coupled parameters do not show any noticeable variability with pulse amplitude when they are left free.

\begin{figure}[h]
\center{\includegraphics[width=1\linewidth]{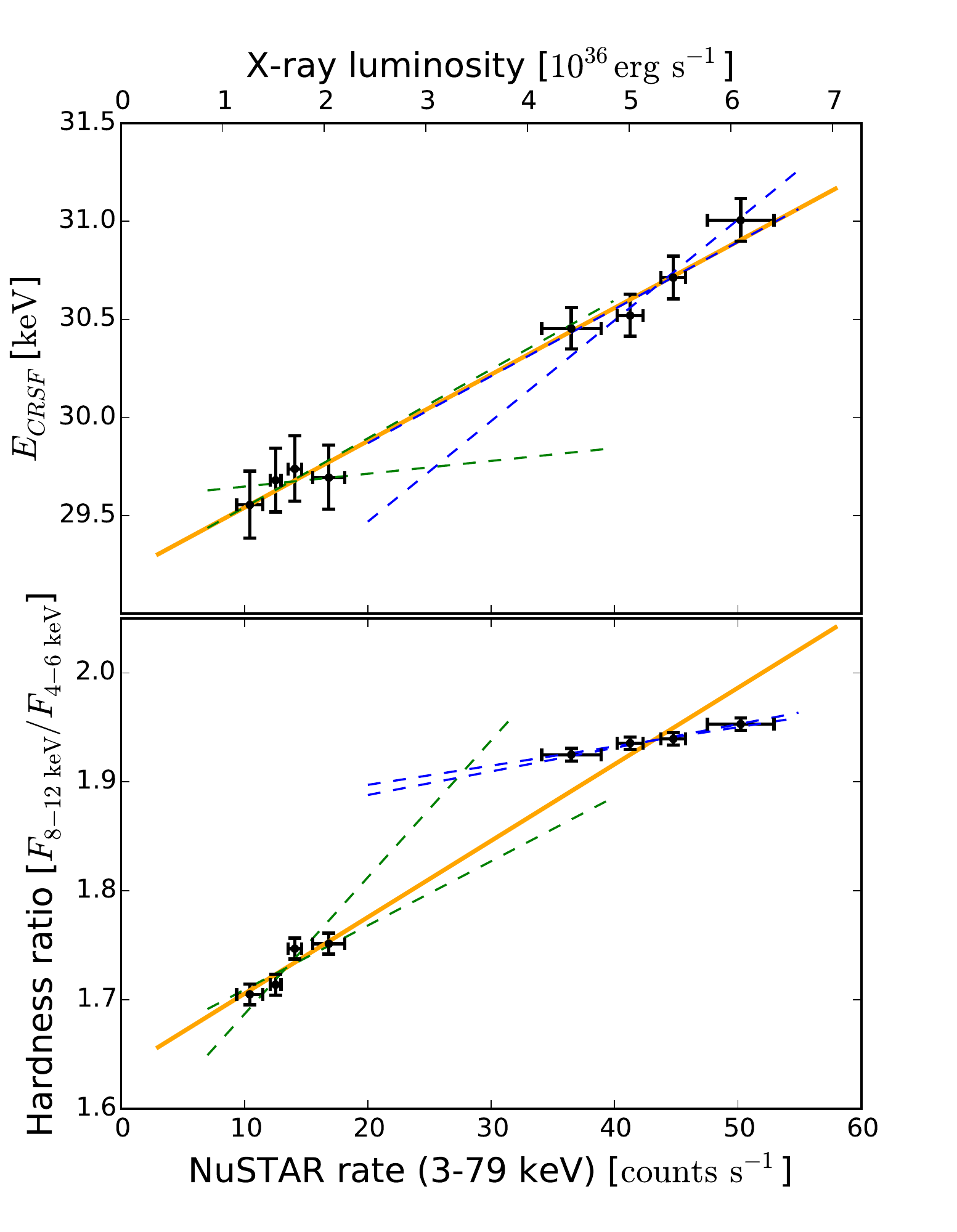}}
\caption{Fundamental CRSF centroid energy (top) and the spectral hardness ratio (bottom) of Cep X-4 as a function of pulse amplitude, which is converted to luminosity on the top axis. The orange solid line represents a common linear fit to the data points for the two observations. The dashed lines mark 1$\sigma$ cones of the allowed slope ranges from the separate fits to the two observations.}
\label{fig:E-H}
\end{figure}

The dependencies of the fundamental CRSF centroid energy and of the continuum hardness ratio on pulse amplitude are shown in Fig.~\ref{fig:E-H}. The bottom axis represents the bin-averaged count rate in the 3-79 keV range, while the top axis represents the corresponding X-ray luminosity in the same energy band for the assumed distance of 3.8\,kpc. The vertical and horizontal error bars on the data points in Fig.~\ref{fig:E-H} correspond to 1$\sigma$ uncertainties.

Since the emission of the pulsar is not isotropic, the flux in the pulse cannot be converted into luminosity by simply multiplying by $4\pi D$, where $D$ is the source distance. Therefore we first calculated the pulse-averaged flux of each observation. This
flux may to some extend be considered as the source flux averaged over the solid angle and can therefore be converted into luminosity in the usual way. Using the pulse averaged flux, we calculated the average luminosity for each of the two observations. We then assumed that the relative changes in the pulse amplitude with respect to the average pulse amplitude in an observation are equal to the relative change in the luminosity with respect to the average luminosity in this observation. Under this assumption, we calculated the luminosity of each pulse or the average luminosity in each pulse amplitude bin, which is represented by the top axis of Fig.~\ref{fig:E-H}.

As shown in the top panel of Fig.~\ref{fig:E-H}, the data reveal a strong positive correlation between the fundamental CRSF energy and the X-ray luminosity. Furthermore, the correlation is present in the data of observation~I alone (data points in the top right), where the differences in luminosity between the data points are caused by the pulse-to-pulse variability. The data of observation~II alone (data points on the bottom left) have insufficient statistics to claim a correlation. 

The common linear fit to all the data points is shown with a solid line and yields a slope of $0.034 \pm 0.002 \,\textrm{keV} \,\textrm{(cts/s)}^{-1}$. This slope is formally consistent with separate linear fits to the data points of the two  observations with slopes of $0.043\pm 0.009$ and of $0.021\pm 0.014 \,\textrm{keV} \,\textrm{(cts/s)}^{-1}$ for observations I and II, respectively. The dashed lines mark 1$\sigma$ cones of the allowed slope ranges from the separate fits of the two observations. It is therefore very likely that the $E_\text{CRSF}/L_X$ dependence is roughly the same in the two observations and between them.

The bottom panel of Fig.~\ref{fig:E-H} shows the spectral hardness as a function of pulse amplitude or luminosity. 
The hardness ratio clearly rises with increasing luminosity from observation~II to observation~I as well as within each observation. It can also be seen that the dependence is substantially flatter in observation~I. Linear fits have slopes of $(2.0\pm 0.2) \times 10^{-3}$ and of $(9.2\pm 3.3) \times 10^{-3} \,\textrm{(cts/s)}^{-1}$ for observations I and II, respectively (dashed lines). The difference in slopes is thus statistically significant.

To ensure the reliability of the dependencies, we fitted the spectra with alternative phenomenological models. Specifically, we used a power law with an exponential cut-off 
\begin{equation}
\textrm{highecut(E)} \propto 
  \begin{cases}
     E^{-\Gamma} & E \le E_{\textrm{cut}} \\
     E^{-\Gamma}\exp\left(-\frac{E-E_{\textrm{cut}}}{E_{\textrm{fold}}}\right) & E > E_{\textrm{cut}}
  \end{cases}
\end{equation}
\citepads[\texttt{highecut} in XSPEC, ]{1983ApJ...270..711W} 
and an additive emission Gaussian line to flatten the 10~keV feature. We also added a \texttt{gabs} component in the continuum to smooth the derivative discontinuity at $E_{\textrm{cut}}$ in this case.
The CRSF was also described as a multiplicative absorption line with a Lorentzian optical depth profile (\texttt{lorentz} in XSPEC). 
In addition to our model, we thus tested three alternative models: the \texttt{highecut*gabs+gauss} model for the continuum with either the \texttt{gabs} or the \texttt{lorentz} model for the cyclotron line, and the \texttt{fdcut+gauss} model for the continuum with the \texttt{lorentz} model for the cyclotron line. The iron line was described as an additive emission \texttt{gauss} in all models. 
All the dependencies described above are found to be present in all the alternative models, and all our conclusions remain valid independently of the choice of a specific spectral function.

\section{Discussion}
\label{sec:disc}
\subsection{Timescales of the correlations}
The aforementioned dependencies between the fundamental CRSF energy, the spectral hardness ratio, and the X-ray luminosity in separate observations are results from the pulse-to-pulse analysis, which makes use of an intrinsic source pulse-to-pulse variability. In other words, the dependencies for separate observations are present on a timescale corresponding to one cycle of X-ray pulsations, that is, on the order of a minute. 

In the case of the $E_{\textrm{CRSF}}/L_\textrm{x}$ correlation, however, the same linear dependence is likely to be present on both short (a pulse cycle) and long (outburst duration) timescales. We can therefore assume that the physical processes behind the positive $E_{\textrm{CRSF}}/L_\textrm{x}$ dependence observed in Cep X-4 are linear, in a first approximation, on timescales from a minute to a dozen days, or in other words, the properties of these processes remain constant on these timescales.

In the case of the spectral hardness, on the other hand, such a linearity is not observed, since the linear dependencies in separate observations are not consistent with the common linear dependence. Thus, the shape of the $HR/L_\textrm{x}$ dependence is likely to be more complex on long timescales of days, while on short timescales it seems to be linear.

\subsection{Accretion regimes}
As mentioned in Sect. 1, positive and negative correlations between the CRSF energy $E_{\textrm{CRSF}}$ and the X-ray luminosity $L_\textrm{x}$ are observed in some X-ray pulsars. 
A certain threshold or critical luminosity $L_{\textrm{cr}}\sim 10^{37}$\,erg~s$^{-1}$ that reflects the mass accretion rate $\dot{M}$ is believed to separate the accretion modes that are
characterised by different signs of the correlation \citepads{1976MNRAS.175..395B, 2012A&A...544A.123B, 2015MNRAS.447.1847M}. A positive $E_{\textrm{CRSF}}/L_\textrm{x}$ correlation is observed below $L_{\textrm{cr}}$ , while a negative correlation is observed above it.

Above $L_{\textrm{cr}}$, the radiation pressure in the polar emitting region of the NS dominates gas pressure such that the infalling plasma is decelerated mainly through interactions of the plasma electrons with photons. This mode is often referred to as supercritical. The hight of the emitting region, which in this case is believed to have roughly a column-like shape (accretion column), is expected to grow with mass accretion rate \citepads{1973NPhS..246....1D, 1976MNRAS.175..395B,1981A&A....93..255W, 1991ApJ...367..575B, 2015MNRAS.452.1601P, 2015MNRAS.454.2539M}.
If the CRSF is formed in a region associated with the accretion column as suggested in \citetads{2012A&A...544A.123B}, for example,
the cyclotron line energy will be anti-correlated with the height of the column. Indeed, the strength of the NS dipole magnetic field decreases with increasing height above the stellar surface and so does the cyclotron line energy, which is proportional to the field strength.

Alternatively, the CRSF might form in the radiation reflected by the NS surface surrounding the accretion column, as was suggested by \citetads{2013ApJ...777..115P}. A growth of the column would widen the stellar surface that is illuminated by the radiation from the column, including the areas farther away from the magnetic pole where magnetic field is weaker. As a result, an anti-correlation between the line energy and the accretion rate should be observed.

In accreting pulsars with luminosities below $L_{\textrm{cr}}$, which are sometimes referred to as subcritical sources, the radiation pressure is unimportant compared to gas pressure. In this case, the infalling plasma has to be decelerated through
Coulomb collisions, collective plasma effects, or through a collisionless shock that forms above the stellar surface.
The following ideas have been proposed in the literature to explain the positive $E_{\textrm{CRSF}}/L_\textrm{x}$ correlation in the subcritical case. 
In \citetads{2007A&A...465L..25S} and \citetads{2012A&A...544A.123B}, the line-forming region was again assumed to be associated with the stopping region such that variations in the height of this region above the stellar surface lead to corresponding changes in the CRSF energy. It was shown that contrary to the supercritical case, the stopping region approaches the surface with increasing mass accretion rate. Therefore an anti-correlation between the CRSF centroid energy and luminosity is expected. 
An alternative idea has been elaborated on in \citetads{2015MNRAS.454.2714M}, who suggested that the variations in the energy of the cyclotron line are due to the Doppler effect. The dependence of the mean velocity of the infalling matter in the line-forming region on the mass accretion rate is responsible for the positive
$E_{\textrm{CRSF}}/L_\textrm{x}$ correlation in this model.

\subsection{Colissionless shock model}
Here, we consider a so far poorly explored possibility that infalling matter decelerates in a collisionless shock forming above the NS surface \citepads{1982ApJ...257..733L}. 
This is likely to be the main stopping agent of the accretion flow when the photon luminosities are low enough for radiation pressure to be unimportant for the accretion flow breaking (but in the CRSF, where the optical depth is very high
even at low luminosities, see \citetads{2004AstL...30..309B}).
This model has recently been used to describe the observed non-linear dependencies of the CRSF parameters on luminosity in another subcritical X-ray pulsar GX\,304$-$1 \citepads{2017MNRAS.466.2752R}. We develop this model in more detail so that it can also be used to describe the evolution of the spectral hardness ratio (HR) with luminosity. 
 
\subsubsection{CRSF energy -- luminosity dependence}
The height of the collisionless shock forming in the infalling flow near the NS surface is expected to depend on electron density $n_{\textrm{e}}$ as  $H_\textrm{s} \propto 1/n_{\textrm{e}}$ \citepads{1975ApJ...198..671S}. The electron density $n_\textrm{e} \propto \dot{M}/A$, where $\dot{M}$ is the mass accretion rate and $A$ is the accretion area of the accretion flow on the stellar surface depending on the magnetospheric radius $R_\textrm{m}$ as $1/R_\textrm{m}$ in a dipole field approximation. The magnetospheric radius $R_m \propto \dot{M}^{-x}$, where $x = 2/7$ for disc accretion or Bondi quasi-spherical accretion, and $x = 2/11$ for quasi-spherical settling accretion \citepads{2012MNRAS.420..216S}, so that $A \propto \dot{M}^x$ and
\begin{equation}\label{e:Hs-M}
H_\textrm{s} \propto 1/n_\textrm{e} \propto A/\dot{M} \propto \dot{M}^{x-1} \propto \dot{M}^{-\alpha},
\end{equation}
where $\alpha = 1-x = 5/7$ for disc accretion or Bondi quasi-spherical accretion, and $\alpha = 9/11$ for quasi-spherical settling accretion. Assuming the disc accretion in Cep X-4, we use $\alpha = 5/7$ in all formulas below.

\begin{figure}[h]
\center{\includegraphics[width=1\linewidth]{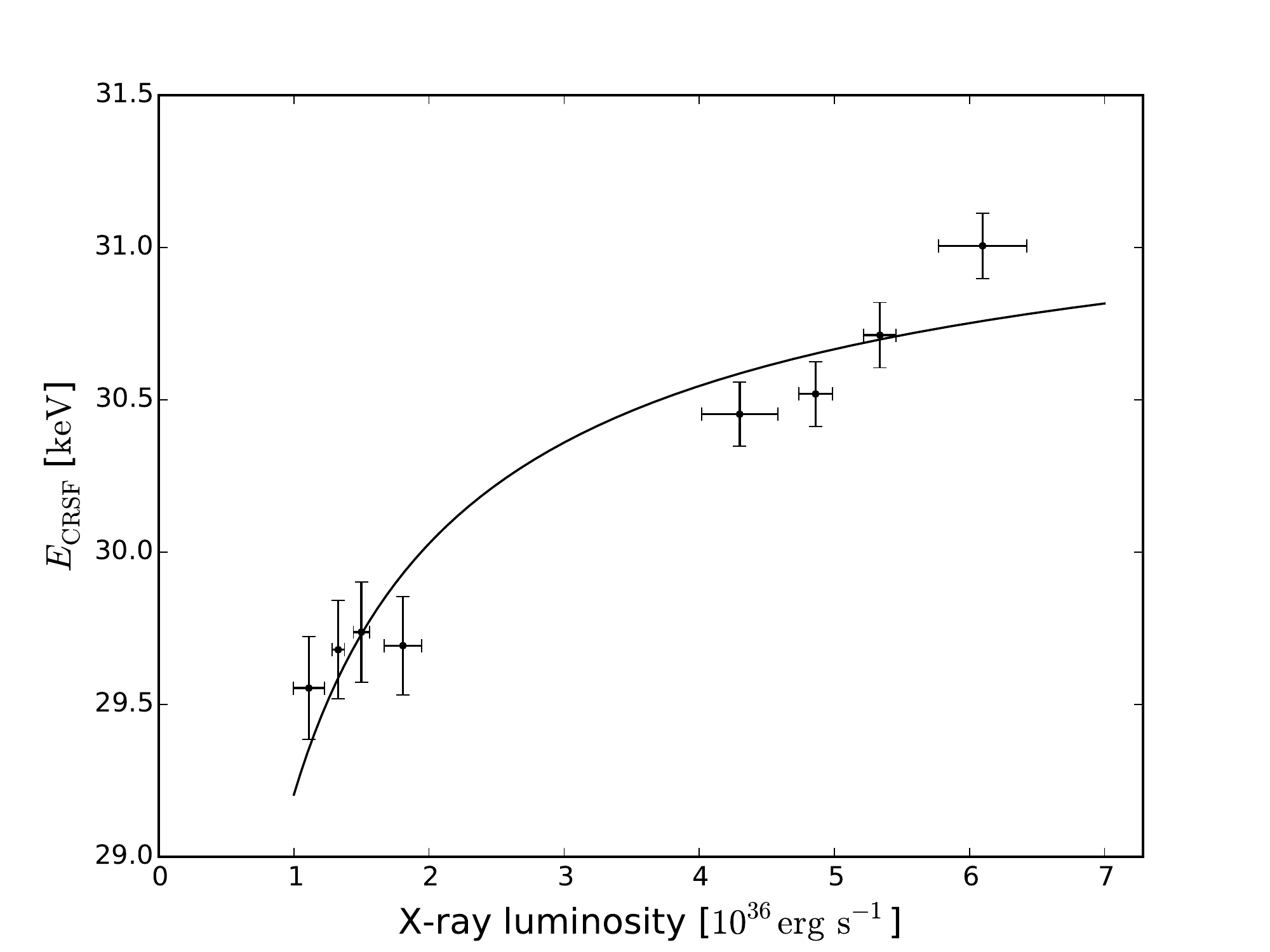}}
\caption{Centroid energy of the fundamental cyclotron line in Cep X-4 as a function of X-ray luminosity (data points with 1$\sigma$
error bars). The solid curve represents a fit using our model assuming a collisionless shock (see text), $\chi^2_{\textrm{red}}/\textrm{d.o.f.} = 1.75/7$. }
\label{fig:E_CepX4-GX304_1}
\end{figure}

With increasing mass accreting rate, the height of the collisionless shock thus decreases such that the line-forming region moves towards stronger magnetic fields and the cyclotron line shifts to higher energies. 
In a dipole field, $E_{\textrm{CRSF}}(r)\propto B(r) \propto 1/r^{3}$ , and considering that $r = R_{\textrm{NS}}+H_\textrm{s}$, where $R_{\textrm{NS}}$ is the NS radius and $H_\textrm{s}$ is a function of accretion rate, we express the energy of the cyclotron line as a function of mass accretion rate:
\begin{equation}
E_{\textrm{CRSF}}(\dot{M}) = E_\textrm{0}\left(\frac{R_{\textrm{NS}}}{H_\textrm{s}(\dot{M}) + R_{\textrm{NS}}}\right)^3, 
\end{equation}
where $E_\textrm{0}$ is the cyclotron line energy corresponding to the magnetic field at the NS surface on the magnetic poles.
Assuming that the observed X-ray flux $F_\textrm{x}$ is directly proportional to the mass accretion rate $\dot{M}$ and using \Eq{e:Hs-M}, we obtain that $H_\textrm{s} \propto F_\textrm{x}^{-\alpha}$. Introducing the dimensionless quantity $H_\textrm{s}/R_{\textrm{NS}}$, we can write 
\beq{e:Hs}
H_\textrm{s} /R_{\textrm{NS}} = KF_\textrm{x}^{-\alpha}
,\eeq
and finally,
\beq{e:E_crsf}
E_{\textrm{CRSF}}(F_\textrm{x}) = E_\textrm{0}(K F_\textrm{x}^{-\alpha} + 1)^{-3}, 
\eeq
where the factor $K$ determines the height of the collisionless shock. $E_\textrm{0}$ and $K$ are found as fit parameters of observational data. In our case, $E_\textrm{0} = 31.4 \pm 0.2$ keV and $K = 0.024 \pm 0.003 \,\textrm{(erg/s)}^{\alpha}$.
 
As shown in Fig.~\ref{fig:E_CepX4-GX304_1}, the collisionless shock model represented with the solid curve describes the observed positive $E_{\textrm{CRSF}}/L_\textrm{x}$ correlation reasonably well. 
We note the apparent turn-up of CRSF energies at the high end of the explored X-ray luminosity range, however. The model
presented above does not predict such a behaviour, but only yields a flattening at high luminosities. If this were real, 
the visible turn-up may indicate gradual deviations from the model being considered here. Indeed, with increasing 
mass accretion rate, the radiation pressure becomes more and more important, so that the simplest version of the
collisionless shock may not be fully applicable for the brightest states. This intermediate case between the two regimes is the most difficult to treat, and apparently full-radiation hydro simulations are needed to investigate this interval of luminosities. We also note that the limited number of data points and their accuracy do not exclude the possibility that the indication of the second derivative is a pure statistical or systematical effect.

Substituting measured $F_\textrm{x}$, found from \Eq{e:E_crsf} $K$ and assuming $R_{\textrm{NS}}=10 \,\textrm{km}$ in \Eq{e:Hs}, we obtain averaged (over each observation) shock heights of $\sim$200 and 100\,m above the NS surface for observations II (in the outburst decay) and~I (near the outburst maximum), respectively.

\subsubsection{Hardness ratio -- luminosity dependence}
The X-ray spectral analysis presented in Sect.~\ref{sec:spe} revealed a strong and apparently non-linear correlation between the spectral hardness ratio and X-ray luminosity in Cep X-4.
Generally, an increase of spectral hardness with luminosity is expected to be due to the increase of the electron number density $n_e$  with increasing $\dot{M,}$ leading to more effective Comptonisation of the emerging X-ray spectrum. 
In the model with a collisionless shock, most of the energy is released near the NS surface where the continuum spectrum Comptonised by electrons between the NS surface and the shock is produced. Numerical simulations performed by \citetads{2004AstL...30..309B} suggest that at a given accretion rate neither $T_e$ and $n_e$ change significantly downstream of the shock. The shape of the emerging spectrum depends on the energy exchange between photons and electrons (the dimensionless Comptonisation $y$-parameter, \citetads{Kompaneets56}), which in the strong magnetic field is \citepads{1975A&A....42..311B}
\beq{e:y}
y=\frac{2}{15}\frac{kT_e}{m_ec^2}\text{Max}\{\tau, \tau^2\}\,,
\eeq
where $\tau$ is the characteristic optical depth of the region and all other notations are standard. Consider a 
cylindrical geometry of the accretion region with the polar cap area $A$ and height $H_s$. In real X-ray observations, the geometry is averaged over the viewing angle by the rotation of the NS, therefore the appropriate characteristic scale of the problem is    
\beq{e:reff}
r_{eff}=(AH_s)^{1/3}\,.
\eeq
Comptonisation is effective for $y\sim 1$, and the number of scatterings in this region is 
$\sim \tau_{eff}^2\propto n_e^2r_{eff}^2$. The effective $y$-parameter of the problem therefore scales as  
\beq{e:yeff}
y_{eff}\propto\tau_{eff}^2 \propto n_e^2 r_{eff}^2 \propto \left(\frac{\dot M}{A}\right)^2 (AH_s)^{\frac{2}{3}} 
\propto \dot M^\frac{10}{7} \dot M^\frac{4}{21} \dot M^{-\frac{10}{21}}\propto \dot M^\frac{24}{21}\,,
\eeq
where we have made use of the standard dependence $A\propto \dot M^{2/7}$ (a dipole magnetic field geometry). Next, we use the 
standard solution of the unsaturated Comptonisation problem \citepads{1983ASPRv...2..189P}, where the emerging spectral flux
at photon energies $h\nu < kT_e$ can be written as
\beq{e:F_nu}
F_\nu=C\nu^{-\alpha}
\eeq
with the spectral index 
\beq{e:alpha}
\alpha=-\frac{3}{2}+\sqrt{\frac{9}{4}+\frac{2}{15y_{eff}}}\,.
\eeq
(Here we made use of the decrease in the efficiency of the energy exchange between electrons 
and photons in a strong magnetic field, see \Eq{e:y} above). The spectral hardness ratio is determined in the usual way:
\beq{e:HR}
\mathrm{HR}=\frac{\int\limits_{\nu_3}^{\nu_4} F_\nu d\nu}{\int\limits_{\nu_1}^{\nu_2} F_\nu d\nu},
\eeq
where, in our case, the frequencies  $\nu_1$, $\nu_2$, $\nu_3$ , and $\nu_4$ correspond to energies of 4, 6, 8, and 12 keV, respectively.
Plugging \Eq{e:F_nu} and \Eq{e:alpha} into \Eq{e:HR} and taking \Eq{e:yeff} into account, we arrive at
\beq{HR}
\mathrm{HR}=\frac{\int\limits_{\nu_3}^{\nu_4} \nu^{3/2-\sqrt{9/4+{2}/{(15y_{eff})}}} d\nu}
{\int\limits_{\nu_1}^{\nu_2} \nu^{{3}/{2}-\sqrt{{9}/{4}+{2}/{(15y_{eff})}}} d\nu}=
\frac{\int\limits_{\nu_3}^{\nu_4} \nu^{3/2-\sqrt{9/4+2/(15K_y\dot M^{24/21})}} d\nu}
{\int\limits_{\nu_1}^{\nu_2} \nu^{{3}/{2}-\sqrt{{9}/{4}+2/(15K_y\dot M^{24/21})}} d\nu}\,,
\eeq
where the parameter $K_y$ is found from fitting of the observed HR change with X-ray flux.
 
\begin{figure}[h]
\center{\includegraphics[width=1\linewidth]{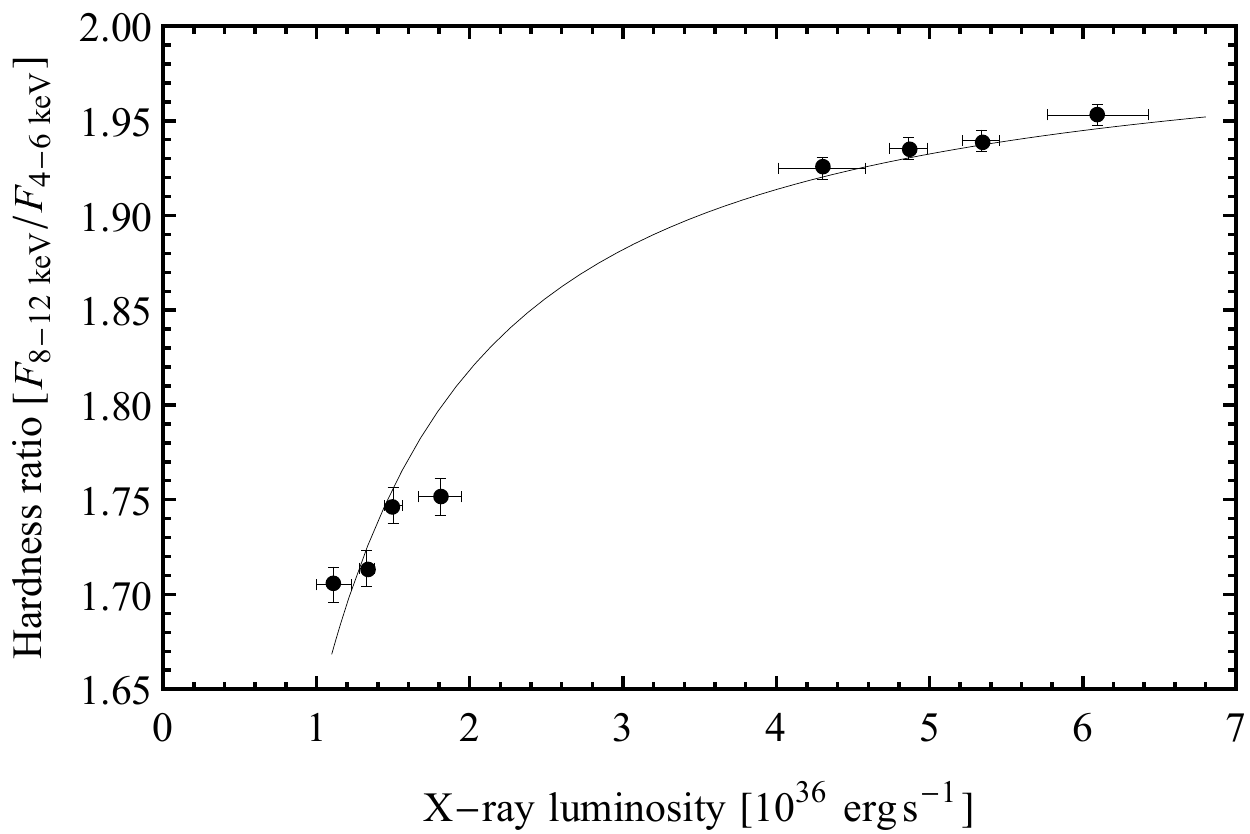}}
\caption{Similar to Fig.~\ref{fig:E_CepX4-GX304_1}, but for the
spectral hardness ratio. The solid curve represents a model expressed by \Eq{HR}, $\chi^2_{\textrm{red}}/\textrm{d.o.f.} = 1.36/7$.}
\label{f:HR_fit}
\end{figure}

The best-fit model is presented in Fig.~\ref{f:HR_fit}. Clearly, the model of \Eq{HR} describes the observed behaviour of hardness ratio with increasing X-ray luminosity of Cep X-4 well. 
The best-fit value of the sole parameter is $K_y\simeq 0.14$, which corresponds to an effective $y$-parameter
range $\sim 0.2-1.2$ during the observed luminosity change in Cep X-4. 
These values of $y_{eff}$ are consistent with the unsaturated Comptonisation regime, which 
justifies the use of \Eq{e:F_nu} and \Eq{e:alpha} for the model fitting. 

To illustrate the qualitative analysis given above, we performed numerical simulations of radiative transfer in the scattering medium above the neutron star polar cap using a two-dimensional Feautrier method to obtain the hardness ratio. The details of the numerical calculations are presented in Appendix~A.

\begin{figure}[h]
\center{\includegraphics[width=1\linewidth]{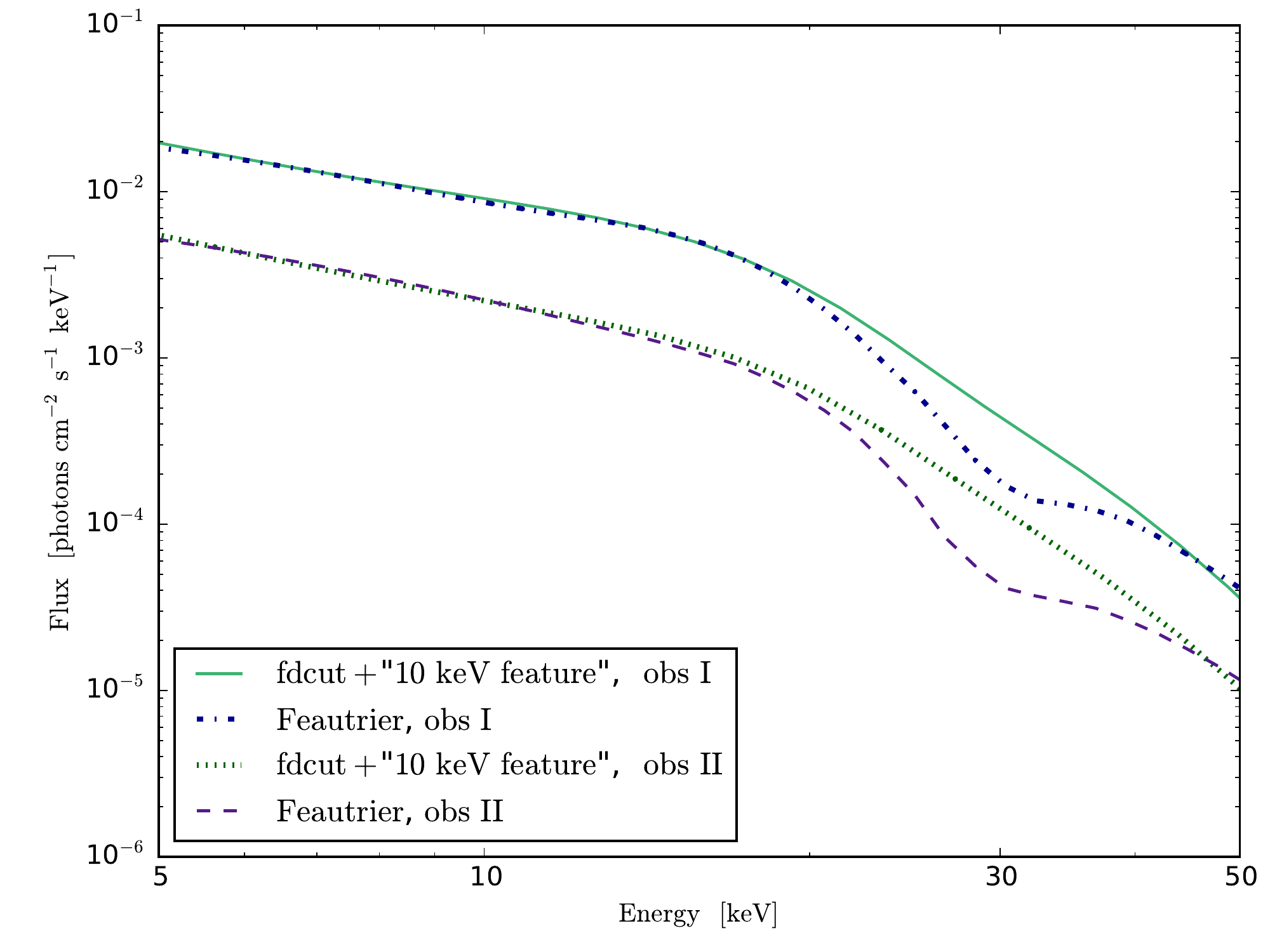}}
\caption{Model spectra with the observed \texttt{fdcut} continua for observations~I and~II. 
(I): $\dot{M}=5\cdot10^{16}$ g s$^{-1}$, the polar cap radius $r_{0}=0.38$ km, the height of the column $z_{0}=77$ m, $n_{e}=2.66\cdot10^{21}$ cm$^{-3}$. 
(II): $\dot{M}=1.5\cdot10^{16}$ g s$^{-1}$,  $r_{0}=0.32$  km, $z_{0}=180$ m, electron density $n_{e}=1.12\cdot10^{21}$ cm$^{-3}$. 
The temperature of the scattering electrons is $T_{e}=5$ keV in both cases.}
\label{fig:num-spectra}
\end{figure}

Figure~\ref{fig:num-spectra} shows the emerging model spectra from the accretion column calculated as specific flux per unit surface integrated over the entire column for observations~I (dashed line) and~II (dash-dotted line). To compare them with the observed continua (solid and dotted lines), the modelled specific luminosity from two columns was divided by some factor to match the continuum given by Eq.(1). This factor is equal to $4\pi d^2$, where $d \sim 4~\mathrm{kpc}$ is the 
distance to the source.

The hardness ratios of the modelled spectra are calculated to be 1.93 and 1.74 for observations~I and~II, respectively, which is very similar to the observed values (see Fig.~\ref{fig:E-H}).
This shows that the model spectra are in good agreement with the observed continua, confirming the expected hardening of the continuum with increasing mass accretion rate. It should be noted that here we did not study the $E_{CRSF}/L_{x}$ correlation obtained from the simulations because the accuracy of the calculations is not high enough to analyse the line centroid energy. The application of the Feautrier method for the cyclotron line energy variations will be explored in the future work.

We therefore conclude that changing the electron density downstream of the collisionless shock 
above the NS polar cap is consistent with the observed hardness ratio evolution with X-ray luminosity in Cep X-4.

\section{Summary}
We have investigated the spectral-luminosity variability in the X-ray pulsar Cep X-4 using NuSTAR observations of the 2014 outburst. The intrinsic pulse-to-pulse variability of Cep X-4 allowed us to reveal strong positive $E_{\textrm{CRSF}}/L_\textrm{x}$ correlations on short timescales of one cycle of X-ray pulsations (minutes) as well as on long timescales of weeks (at least 12 days during which the source becomes fainter by a factor of about 3.5). Moreover, linear fits to the data points seem to have the same slope, which might suggest linearity of the processes behind the $E_{\textrm{CRSF}}/L_\textrm{x}$ correlations.

We also confirm that the continuum becomes harder with X-ray luminosity, that is, the slope of the power law decreases, or in other words, the spectral hardness ratio ($F_{8-12\,\mathrm{keV}}/F_{4-6\,\mathrm{keV}}$) increases with X-ray flux. However, the slope of the linear fit to the points of the measured hardness ratio clearly changes from the first observation to the second, which implies a more complex form of the dependence than a linear one on long timescales (during at least one outburst).

The evolution of the spectral hardness ratio with X-ray luminosity is successfully described by a model in which a collisionless shock above the neutron star surface is formed. This model also describes the observed positive $E_{\textrm{CRSF}}/L_\textrm{x}$ correlation reasonably well. Within this model, heights of cyclotron line emerging layers were estimated. 

Furthermore, we found a harmonic of the cyclotron feature in both NuSTAR observations. We thus confirm the harmonic of the cyclotron line in the Cep X-4 spectrum, as previously found by \citetads{2015MNRAS.453L..21J} in Suzaku data.

\begin{acknowledgements}
The research is supported by the joint DFG grant KL~2734/2-1 and Wi~1860~11-1 and the RFBR grant 14-02-91345. 
We thank the anonymous referee for the useful comments and suggestions that substantially improved the manuscript. The research used the data obtained from HEASARC Online Service provided by the NASA/GSFC.
\end{acknowledgements}

\bibliographystyle{aa}
\bibliography{ref}

\newpage

\begin{appendix}
\section{Numerical modelling}\label{ap_a}
The Feautrier method is a powerful tool of radiation transfer calculations in regions with 
given structure (temperature and density distributions) and is widely used for spectrum calculations \citepads{1981ApJ...251..288N, 1981ApJ...251..278N, 1985ApJ...298..147M, 1989ApJ...342..928A}. In these works a one-dimensional model for two particular geometries was applied. 
In a collisionless shock model, where the radiation from the top of the column could be comparable to that of the sidewalls, it is necessary to take a two-dimensional radiative transfer
into account. 

For this reason, we used a generalised two-dimensional Feautrier method \citepads{1970ApJ...161..255C} to calculate the continuum and (with less accuracy) the fundamental CRSF feature. The calculation domain 
is taken to be a cylindrical column filled with plasma of constant temperature and density. The radiative transfer for two polarisation modes is used in the geometrical optics limit. We employ cylindrical coordinates, assuming axial symmetry; $z = 0$ at the NS surface and $r = 0$ at the centre of the column. 

The radiative transfer equation for the scattering atmosphere reads
\beq{e:rtrans}
 \frac{\sin^{2} \theta_{i}}{2\sigma_{i}} \left( \frac{\partial^{2} u_{i}}{\partial r^{2}}+\frac{1}{r} \frac{\partial u_{i}}{\partial r}\right) + \frac{\cos^{2} \theta_{i}}{\sigma_{i}}\frac{\partial^{2} u_{i}}{\partial z^{2}} - \sigma_{i}u_{i} + \sum\limits_{j=1}^N S_{ij}u_{j} = 0.
\end{equation}
Here $\theta$ is the angular coordinate, $u_{i}(\omega, \theta, r, z)=\frac{1}{2}\left[ I_{i}(\omega, \theta, r, z) - I_{i}(\omega, -\theta, r, z)\right]$, where $I$ is the intensity; $\sigma_{i}=\sigma_{i}(\omega, \theta, z)$ is the scattering coefficient independent of the $r$-coordinate because the magnetic field is assumed to be oriented along the $z$-axis.

The redistribution matrix elements $S_{ij}$ are calculated using the equations for differential scattering cross sections from \citetads{1986PhRvD..34..440B} derived in a fully relativistic approximation. Thus the second-order Doppler effect and the non-harmonicity of the electron Landau levels are included in our calculations. We restricted ourselves to considering the $n =1$ Landau level for the intermediate excited state and did not take spin-flip transitions into account.
The derivation of the fully relativistic normal modes including vacuum and plasma effects requires special treatment, and here we used the non-relativistic vacuum-plasma normal modes from \citetads{1985ApJ...298..147M}, for the sake of simplicity.  

The calculations were performed assuming the mass accretion rates $\dot{M}=5 \cdot 10^{16}$ g s$^{-1}$ and $\dot{M}=1.5 \cdot 10^{16}$ g s$^{-1}$ , corresponding to the averaged luminosities of observations I and II, respectively. 
The electron temperature of the column was set to 5.8 keV for both accretion rates, which is close to the electron temperature obtained in numerical modelling \citepads{2004AstL...30..309B}. The electron number density is a function of $\dot{M}$ and the polar cap radius (\Eq{e:Hs-M}).\par
The bottom of the column (polar cap with area $A$ where most of the accretion power is released) 
serves as the source of photons. Because scattering is important in a thin emission region above the polar cap, we assume a modified black-body spectrum of seed photons.
Other boundary conditions are that there is no incident radiation onto the top and the sidewall of the column, $\partial u / \partial r = 0$ along the $z$-axis.

The temperature of the seed radiation is a function of the accretion rate. Taking into account that $L \propto T^{4} A \propto \dot{M}$ and using the standard dependence for the polar cap area, $A \propto \dot{M}^{2/7}$, the temperature ratio for different mass accretion rates is 
\beq{e:trel}
\frac{T_{1}}{T_{2}} = \left( \frac{\dot{M_{1}}}{\dot{M_{2}}} \right)^{5/28} \,.
\end{equation}
The temperature of the radiation coming from the polar cap was fixed to 12 keV for observation~II with the lower mass accretion rate.\par

The results of the simulation for the two accretion rates are shown in Fig. ~\ref{fig:num-spectra} in the main text.

\end{appendix}

\end{document}